\documentstyle[pra,aps,preprint,epsf]{revtex}

\begin{document}
%\draft
\title{Polarization state of a biphoton: \\
quantum ternary logic}
\author{A.~V.~Burlakov, M.~V.~Chekhova, O.~V.~Karabutova,\\ D.~N.~Klyshko,
and S.~P.~Kulik} \address{Department of Physics\\ Moscow State
University\\ 119899 Moscow, Russia}

\maketitle

\begin{abstract}
Polarization state of biphoton light generated via collinear 
frequency-degenerate spontaneous parametric down-conversion is considered. 
A  biphoton is described by a three-component polarization 
vector, its arbitrary transformations relating to the SU(3) group. 
A subset of such transformations, available with retardation plates, 
is realized experimentally. In particular, two independent orthogonally 
polarized beams of type-I biphotons are transformed into a beam of type-II 
biphotons.  Polarized biphotons are suggested as ternary analogs of two-state 
quantum systems (qubits).  \end{abstract}


Polarization state of a single photon 
is described by a two-dimensional
normalized polarization vector. As any quantum
system with two basic states~\cite{Chuang}, an arbitrarily polarized photon can
represent a qubit - a quantum bit of information~\cite{Qubits} used in
quantum computation.
Several quantum logical operations with photon qubits have been
proposed, which make use, in addition to photons, of atoms
or ions~\cite{Atoms}. Recently, quantum
gates were suggested based only on photons,
some of them serving as polarization
qubits and some as location qubits~\cite{Kwiat}.

In this paper, we consider a quantum system formed by two 
correlated photons - a biphoton emitted via
frequency-degenerate collinear spontaneous parametric
down-conversion (SPDC). Its polarization state
is assumed to be arbitrary. In this
general case, the biphoton can be described by the state
vector~\cite{BurKl}

\begin{equation}
\Psi=c_1|2,0\rangle+c_2|1,1\rangle+c_3|0,2\rangle,
\label{1}
\end{equation}
where $c_i=d_ie^{i\phi_i}$
are complex amplitudes and the notation $|N_x,N_y\rangle$ means
a state with $N_x$ photons in the horizontal ($x$) polarization mode and
$N_y$ photons in the vertical ($y$) polarization mode, with $N_x+N_y=2$. The
normalization condition is $\sum_i |c_i|^2=1$. In most cases, the total phase 
of the state (\ref{1}) is not essential, so one can 
assume $\phi_2=0$, and the three-component state of a biphoton is given by 
four real parameters. One can introduce the `polarization vector' of a 
biphoton,

\begin{equation}
{\bf e}=(c_1, c_2, c_3).
\end{equation}

In the most general form, the state (\ref{1}) can be prepared via
SPDC generated in three nonlinear crystals with common coherent
pumping. The states $|2,0\rangle$ and $|0,2\rangle$ are generated via
type-I SPDC, and the state $|1,1\rangle$ via type-II SPDC~\cite{Vacuum}.
According to Eq.(\ref{1}), a biphoton is a three-state system,
similarly to a particle with spin 1. Arbitrary transformations of 
polarization vectors ${\bf e}$ are given by unitary $3\times3$
matrices $G$, $G^+G=I,
\hbox {det}G=1$, which form a three-dimensional
representation of the SU(3) group, see~\cite{Sudbery}.
This type of symmetry, which is well-known in nuclear physics
but seems to be new for optics, could be used for developing
`ternary logic' in quantum computation. To each of the three basic
states $|2,0\rangle$, $|1,1\rangle$, and $|0,2\rangle$, one
can assign one of the digits 0, 1, and 2. The advantage of `ternary'
quantum logic over binary logic (qubits) is the larger number of
states that can be covered by an $n$-element quantum register: $3^n$
instead of $2^n$. The first question
arising here is how one can `switch' between these three basic
states or their combinations. 

According to the properties of the SU(3) group,
an arbitrary transformation $G$ of the vector ${\bf e}$ is given
by eight real parameters.
Linear lossless
elements (retardation plates and polarization rotators) introduced
into the biphoton beam transform the vector ${\bf e}$ but cannot give all
possible matrices $G$. A transformation of this kind can be characterized by
three independent parameters and therefore, it only realises a
three-dimensional representation of the SU(2) group, which leaves invariant
the polarization degree $P$~\cite{BurKl,KlJETP,G}. At the 
same time, two of the basic states in superposition~(\ref{1}) have $P=1$ and 
one has $P=0$.

However, by passing from the basis
${|2,0\rangle, |1,1\rangle, |0,2\rangle}$
 to the basis
\begin{eqnarray}
\nonumber
\Psi_+&=&\frac{|2,0\rangle+|0,2\rangle}{\sqrt{2}}\equiv|+,-\rangle\\
\Psi_-&=&\frac{|2,0\rangle-|0,2\rangle}{\sqrt{2}}\equiv|+45^\circ, 
-45^\circ\rangle,\\
\nonumber
\Psi_0&=& |1,1\rangle\equiv|x,y\rangle, 
\end{eqnarray}
one obtains three states that can be transformed one into 
another by means of only retardation plates. Indeed, all three vectors of the 
new basis have $P=0$. They all correspond to pairs of 
correlated photons with orthogonal polarizations: right and left circular, 
linear at $\pm45\circ$ to x, and along x and y. In this work, we 
experimentally realize transformations between these states.

The states $\Psi_+, \Psi_-, \Psi_0$ have much in common with quantum ternary 
logic states (`trits') suggested in~\cite{trits}. Indeed, the Bell states 
$\frac{1}{\sqrt{2}}(|H\rangle|H\rangle\pm|V\rangle|V\rangle)$ of~\cite{trits} 
correspond to $\Psi_{\pm}$, the Bell state 
$\frac{1}{\sqrt{2}}(|H\rangle|V\rangle+ |V\rangle|H\rangle)$ corresponds to 
$\Psi_0$, and the Bell state $\frac{1}{\sqrt{2}} 
(|H\rangle|V\rangle-|V\rangle|H\rangle)$ has no sense in the case of 
indistinguishable photons.  It is essential that unlike the states introduced 
in~\cite{trits}, all states considered here relate to a single spatial mode. 
This is an important practical advantage, since it removes the requirement of 
equalizing optical paths, which was necessary in~\cite{trits}.

In experiment (Fig.~1), we use a type-I lithium iodate crystal pumped
by cw He-Cd laser with  wavelength 325 nm and vertical
polarization. The pump is
split into two collinear beams, so that horizontally polarized SPDC with
${\bf e}=(1,0,0)$ is generated
in two spatially separated domains. The pump radiation is suppressed
by the cutoff filter F. After the crystal, the SPDC radiation from
one of the domains is passed through a $\lambda/2$ plate oriented at 45 degrees
to the initial polarization. The corresponding biphoton polarization vector
becomes $(0,1,0)$. Both SPDC beams are then joined together by means of a
polarizing beamsplitter PBS1. In fact, this part of the setup is a
Mach--Zehnder interferometer with the nonpolarizing beamsplitter for the pump 
at the input and a polarizing beamsplitter for biphoton radiation at the 
output. After the interferometer, the state is

\begin{equation}
\Psi=\frac{1}{\sqrt{2}}(|2,0\rangle+\hbox{e}^{i\phi}|0,2\rangle),
\label{2}
\end{equation}
where the phase $\phi$ is varied by means of a piezoelectric
element (PE) shifting the mirror at the input of the beamsplitter.
Preparation of the biphoton state is accomplished by introducing a
retardation plate RP (either a half wave plate or a quarter wave plate)
after the beamsplitter.

A half wave plate with the optic axis oriented at the
angle $\chi$ to the horizontal direction transforms~\cite{BurKl}  the state 
(\ref{2}) into the state of the form (\ref{1}) with 

\begin{equation} 
|c_1|^2=|c_3|^2=\frac{1-\hbox{sin}^2 4\chi \hbox{sin}^2 \frac{\phi}{2}}{2},
|c_2|^2=\hbox{sin}^2 4\chi \hbox{sin}^2 \frac{\phi}{2}.
\label{3}
\end{equation}
At $\phi=\pi$ and $\chi=\frac{\pi}{8}, \frac{3\pi}{8}, \dots$,
$|c_1|=|c_3|=0$, i.e., the state (\ref{2}) is  completely transformed
into the state $|1,1\rangle$. In our notation, this is the transition
$\Psi_-\to\Psi_0$. Note that if $\phi=0$, the state (\ref{2}) is invariant to 
the action of a half wave plate, $\Psi_+\to\Psi_+$. Similarly, for a quarter 
wave plate oriented at the angle $\chi$,

\begin{equation}
|c_2|^2=\hbox{sin}^2 2\chi (\hbox{cos}\frac{\phi}{2}+\hbox{cos} 2\chi 
\hbox{sin}\frac{\phi}{2})^2, 
\label{4} 
\end{equation} 
and the transformation from 
the state (\ref{2}) to the state $|1,1\rangle$ is achieved at $\phi=0$, 
$\chi=\frac{\pi}{4}$. This describes the transition $\Psi_+\to\Psi_0$. 
At the same time, a quarter wave plate with $\chi=\frac{\pi}{4}$ leaves 
$\Psi_-$ invariant. 

Transitions from the states $\Psi_-$ and
$\Psi_+$ to the state $\Psi_0$ can be demonstrated experimentally by
measuring the second-order correlation
function of the final state,
\begin{equation}
G^{(2)}_{xy}\equiv\langle\Psi|E_x^{(-)}E_y^{(-)}E_x^{(+)}E_y^{(+)}
|\Psi\rangle,
\label{5}
\end{equation}
where $E_{x,y}^{(\pm)}$ are field operators for the
modes $x$ and $y$.
Indeed, for a state of the form (\ref{1}),
we have $G_{xy}\sim |c_2|^2$.  The correlation function $G_{xy}$ is measured 
 by means of a polarizing beamsplitter PBS2, two photodetectors, D1 
and D2, and a coincidence cirquit CC  (Fig.~1). The pinhole PH with diameter 
$1$ mm and the interference filter IF with FWHM $\Delta\lambda=10$nm and 
central wavelength $\lambda=650$nm are used for the spatial and frequency 
selection of the SPDC collinear frequency-degenerate radiation.  The 
coincidence counting rate $R_c$,  which is proportional to $G_{xy}$, is 
measured either as a function of the optical path length variation (phase 
$\phi$ variation) introduced by the piezoelectric element or as a function of 
the retardation plate orientation (angle $\chi$ variation).

The experimental dependencies obtained with the half wave plate are shown
in Figures~2 and 3. First, we fix the orientation of the plate,
$\chi=\frac{\pi}{8}$, and measure $R_c$ as a function
of $\phi$, which is determined by the voltage applied to the piezoelectric
element~(Fig.~2). In the minima, the state at the output of the 
interferometer is $\Psi_+$, which stays the same after the half wave plate. 
At the maxima, the interferometer creates the state $\Psi_-$, which is then 
transformed into $\Psi_0$ by the half wave plate. Fixing the phase $\phi$ at 
a maximum ($\phi=\pi$), we measure the dependence of $R_c$ on the half wave 
plate angle $\chi$ (Fig.~3a).  High coincidence counting rate at the maxima 
of this dependence (in comparison with accidental coincidence counting rate, 
which is less than $0.1$ sec$^{-1}$) indicates that the state $|1,1\rangle$ 
is formed.  However, to check that $\Psi_-$ is fully transformed into 
$\Psi_0$, we need to measure the correlation functions $G_{xx}$ and $G_{yy}$, 
which are proportional to $|c_1|^2$ and $|c_3|^2$, respectively. Such 
measurements are performed by introducing an additional block before the 
polarizing beamsplitter PBS2. This block (framed by a dashed line in Fig.~1) 
includes a polarizer selecting $x$ or $y$ polarization and a half wave plate 
rotating the polarization by $\frac{\pi}{4}$. With this block introduced into 
the setup, $R_c$ is proportional to $G_{xx}\sim|c_1|^2$ or 
$G_{yy}\sim|c_3|^2$, depending on the polarizer orientation.  For instance, 
Fig.~3b shows the dependence of $|c_1|^2$ on $\chi$ for the phase $\phi$ 
being the same as for Fig.~3a.  One can see that at the angles $\chi$ where 
maxima of $|c_2|^2$ are observed (Fig.~3a), the amplitude $|c_1|$ (and 
similarly, $|c_3|$) is almost completely suppressed (Fig.~3b).  The 
background coincidence counting rate in Fig.~3b (the visibility of the 
interference pattern is 90\%) can be explained by nonequal losses for the 
states $|2,0\rangle$ and $|0,2\rangle$.

Similarly, to perform the transformation $\Psi_{+}\to\Psi_0$, one
should use a quarter wave plate as RP in Fig.~1. The phase $\phi$ in this
case should be equal to $0$. In  Fig.~4, the dependence of $G_{xy}\sim|c_2|^2$
on $\chi$ at $\phi=0$ is shown. In accordance with Eq.~(\ref{4}), the
period of this dependence is twice
larger than in the case of the half wave plate.

All dependencies shown in Figs.~2-4 demonstrate nonclassical interference 
with high visibility.  If both biphoton states $|2,0\rangle$ and $|0,2\rangle$ 
generated in separate spatial domains are projected onto a single 
polarization direction~\cite{JLett}, one can observe interference in 
coincidences regardless of the delay introduced between the SPDC beams. In 
principle, the crystal inside the interferometer can be replaced by two 
separate crystals, placed at different distances from the beamsplitter. The 
only condition for the interference is that the arms of the interferometer 
should not differ by more than the pump coherence length. This property is 
due to the collinear degenerate phase matching used in our experiment. In a 
similar interference experiment with  
noncollinear SPDC performed previously~\cite{Mandel}, equality of the optical 
path lengths for two crystals was required.~\cite{Farfield} 

Another paradoxical feature of this experiment should be pointed out.
The state $|1,1\rangle$, which is what one calls `a type-II biphoton',
is produced by two independent `type-I biphoton states'
$|2,0\rangle$ and $|0,2\rangle$.
At the same time, the biphoton flux is so low (about hundreds of  $s^{-1}$) 
that biphotons, if considered as `wavepackets' with coherence length 
$l_{coh}=\frac{\lambda^2}{\Delta\lambda}\sim40 \mu$, almost never overlap.
This shows that unlike single photons, biphotons should not be viewed
as independent wavepackets~\cite{BurKl,Postcom}. 

Thus, we have demonstrated switching between the three states $\Psi_-$,
$\Psi_+$, and $\Psi_0$: the transitions $\Psi_-\to\Psi_0, \Psi_+\to\Psi_0$
are performed by half wave and quarter wave retardation plates, respectively.  
Note that the transition $\Psi_-\to\Psi_+$ can be performed by introducing
a $\pi$ phase shift between $x$- and $y$- polarized light, i.e., 
by inserting a half wave plate with the axes parallel to $x,y$
directions. It is worth noting that all these transformations are reversible.  

A remarkable property of retardation plates 
is that they leave invariant the number of biphotons,
i.e., do not split photon pairs. This could be used for developing
`biphoton' communication systems where biphotons propagate along a single
direction, for instance, in an optical fiber, and are transformed by
retardation plates.

This work was supported in part by the Russian Foundation for Basic
Research, grants \#99-02-16419 and \#96-15-96673. We also acknowledge
support of the Russian Federation Integration Program `Basic Optics
and Spectroscopy'.

\newpage
\centerline {Figure Captions}

Fig.~1. The experimental setup. CW radiation of He-Cd laser at $325$ nm is 
fed into a Mach-Zehnder interferometer, so that two coherent pump beams 
excite collinear frequency-degenerate SPDC in different spatial domains of a 
$\hbox{LiIO}_3$  crystal. The cutoff filter F suppresses the pump radiation, 
and the $\lambda/2$ plate rotates polarization of the SPDC light in one of 
the arms by $\pi/2$.  The piezoelectric element PE is used for path length 
(phase $\phi$) variation.  The polarizing beamsplitter PBS1 joins two SPDC 
beams together.  The retardation plate RP, either $\lambda/2$ or $\lambda/4$, 
can be rotated by  angle $\chi$.  The registration part of the setup includes 
the interference filter IF and the pinhole PH selecting SPDC radiation, the  
polarising beamsplitter PBS2, two detectors D1, D2, lenses L1, L2, focusing 
the radiation on the detectors, and the coincidence cirquit CC. The framed 
block including a polarizer P and a $\lambda/2$ plate is introduced for 
measuring $G_{xx}, G_{yy}$; without this block, $G_{xy}$ is measured. 

Fig.~2. Coincidence counting rate $R_c\sim G_{xy}\sim |c_2|^2$ as a function 
of the optical path length variation (phase $\phi$ in Fig.~1). The 
$\lambda/2$ plate after the Mach-Zehnder interferometer is oriented at 
$\pi/8$. Maxima of the dependence correspond to the $\Psi_-$ state formed at 
the output of the interferometer; the half wave plate transforms it into 
$|1,1\rangle$.  In the minima, the state at the output of the interferometer 
is $\Psi_+$, and it is invariant to the action of the half wave plate. 

Fig.~3. Coincidence counting rate corresponding to $G_{xy}$ (a) and to 
$G_{xx}$ (b) as a function of the angle $\chi$ of $\lambda/2$ plate. In the 
lower case, the framed block in Fig.~1 is inserted.  For both dependencies, 
the phase $\phi$ introduced by the piezoelectric element is $\pi$, i.e., the 
state at the output of the interferometer is $\Psi_-$. 

Fig.~4. Coincidence counting rate corresponding to $G_{xy}$ as a function of 
the angle $\chi$ of $\lambda/4$ plate. The phase $\phi$ introduced by 
the piezoelectric element is $0$, i.e., the state at the output of the 
interferometer is $\Psi_+$. In the maxima, the plate transforms it into the 
state $\Psi_0$; in the minima, it leaves it invariant.


\begin{thebibliography}{30}
\bibitem{Vacuum} To be exact, the state of light generated via SPDC is given by 
a superposition of vacuum and two-photon states. However, in many 
experiments, including the one discussed here, the vacuum state gives no 
contribution.
\bibitem{G} The explicit form of $G$ in this case is given in~\cite{BurKl}.
\end{thebibliography}
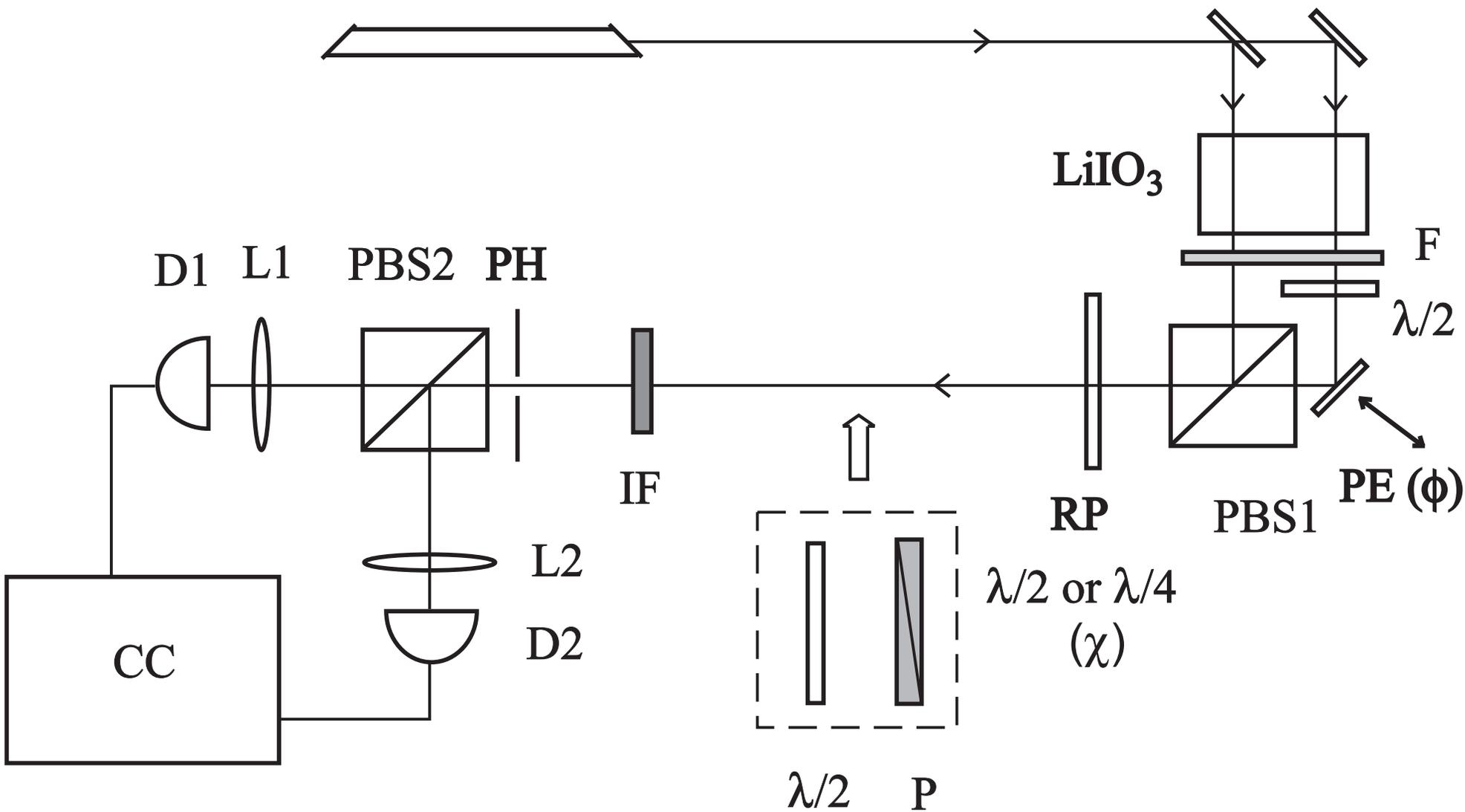
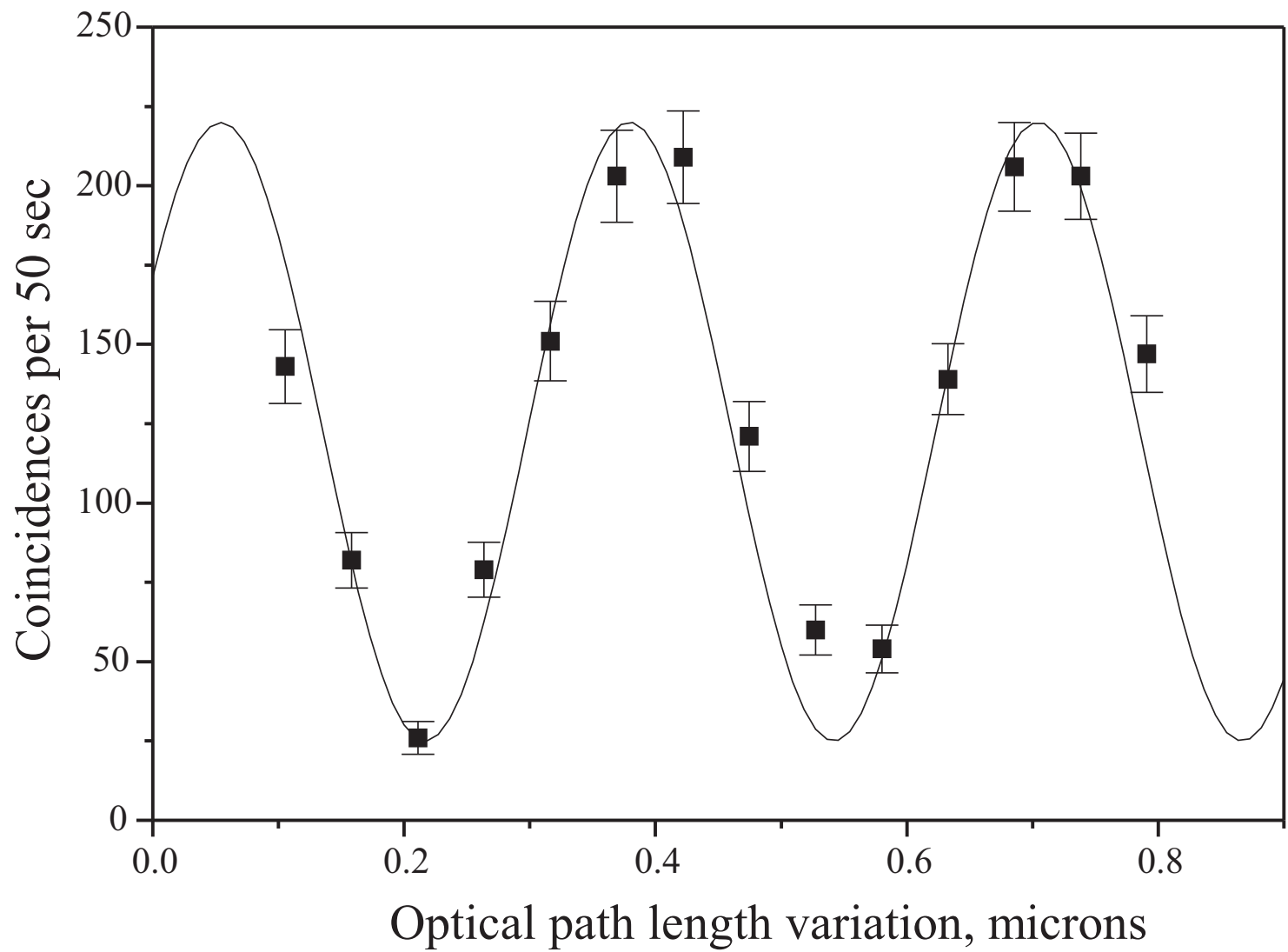
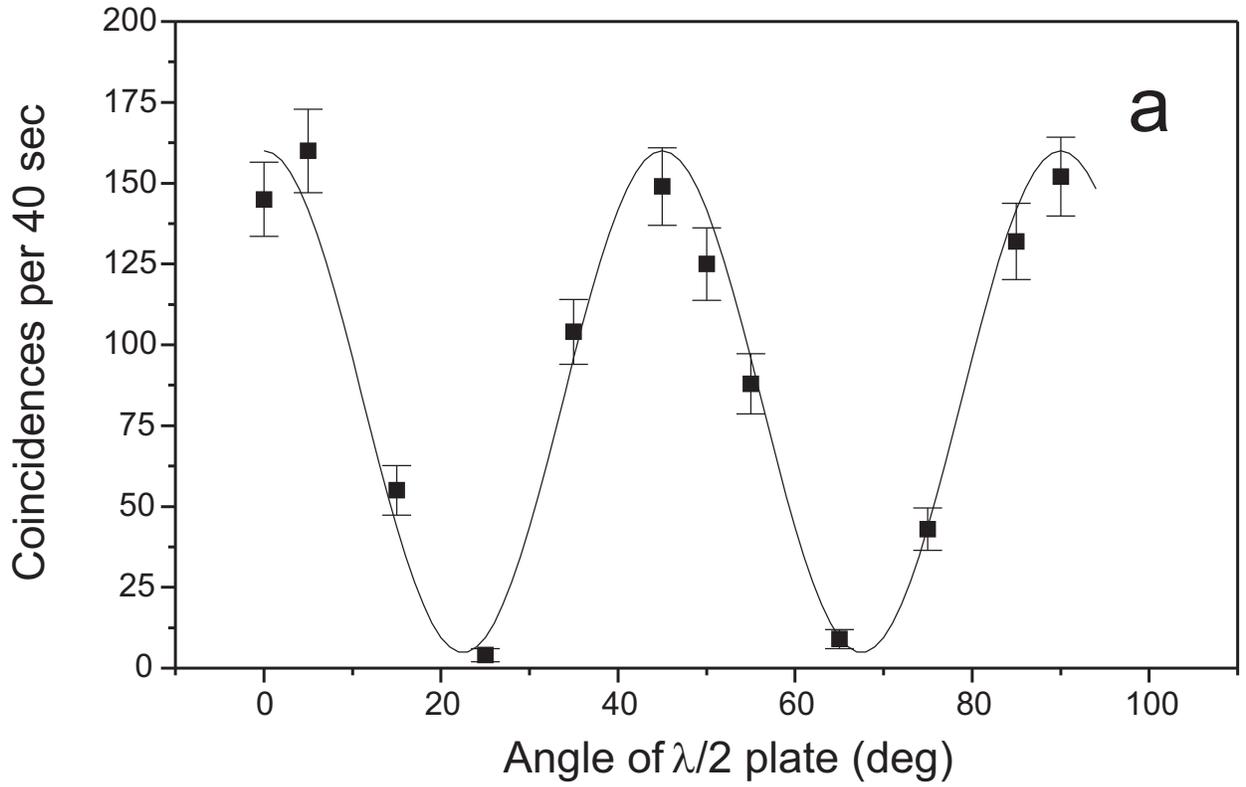
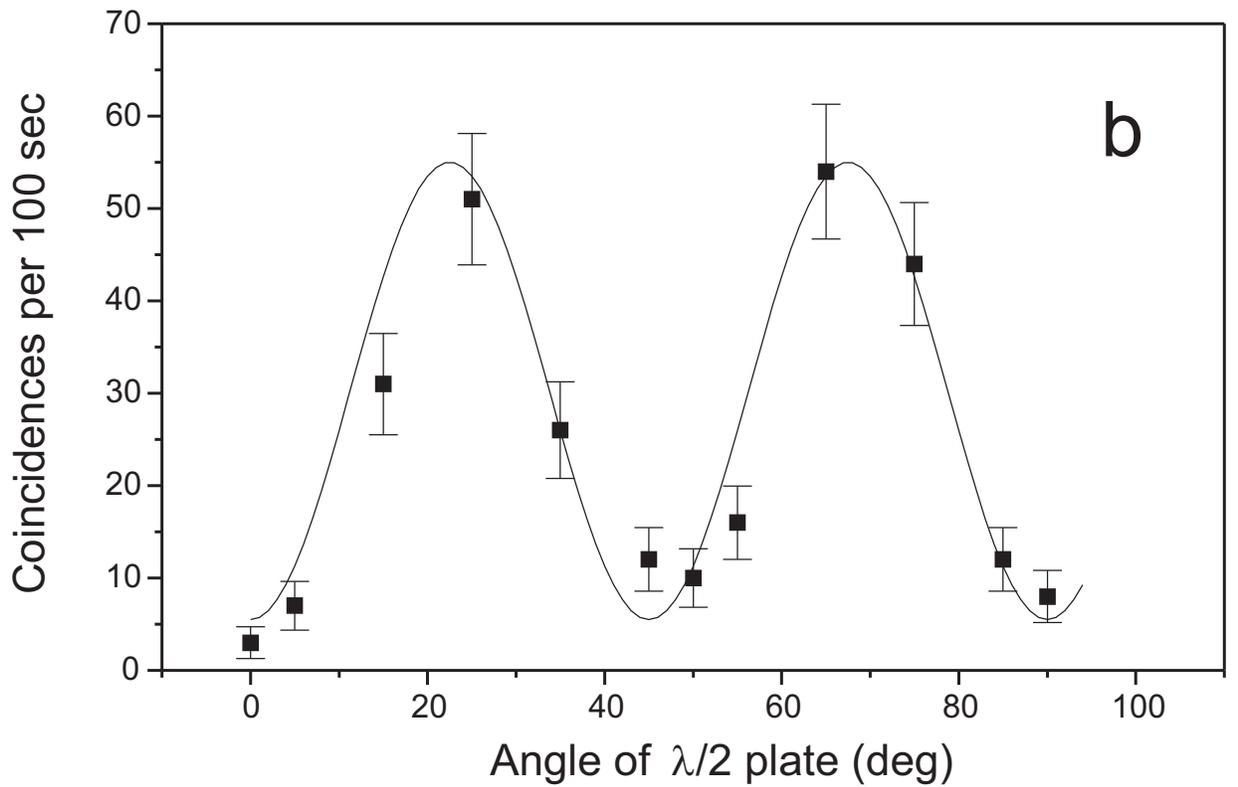
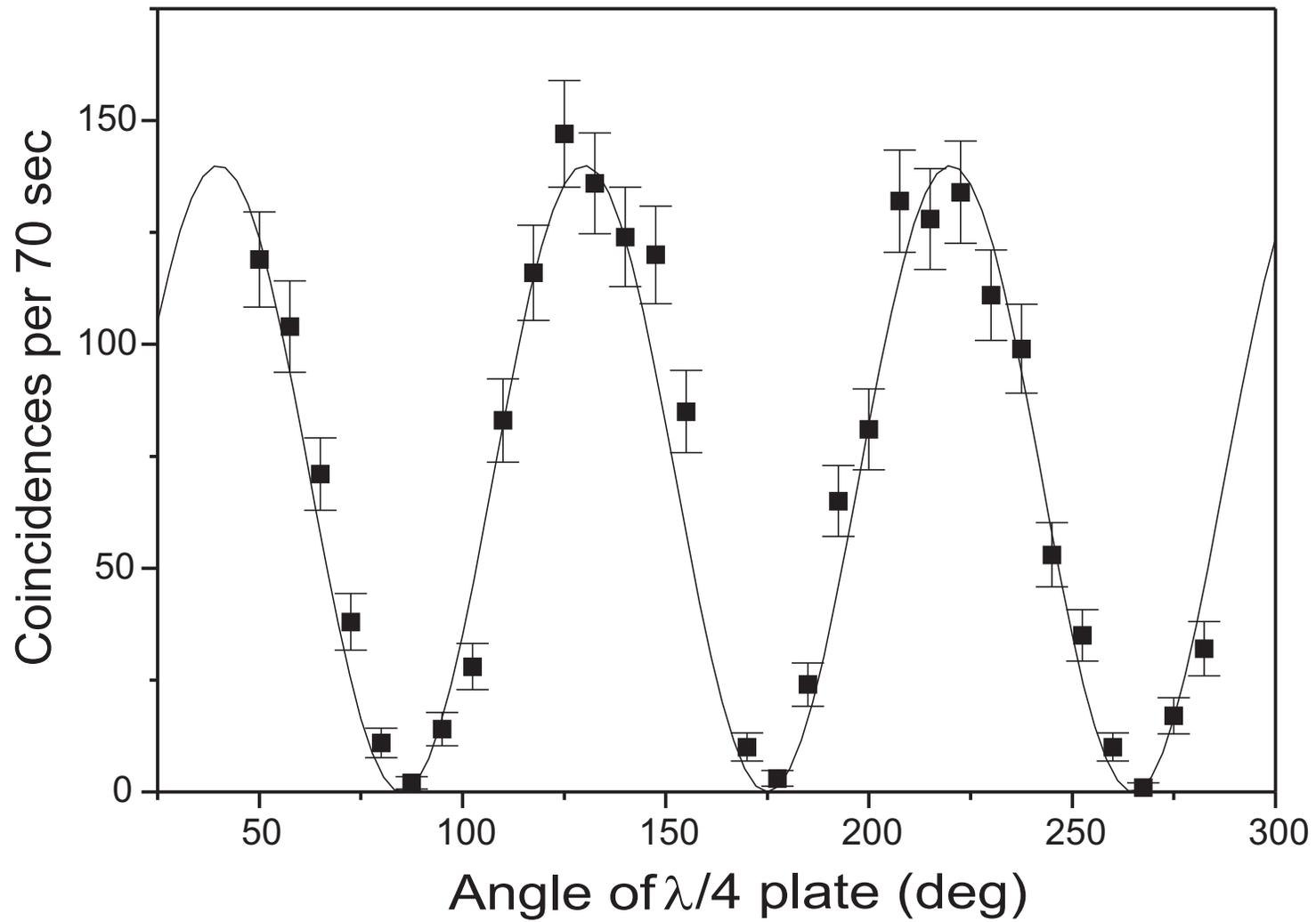
\end{document}